\documentclass{raa}            

\usepackage{graphicx,times}             
\input{epsf.sty}                        
\input{psfig.sty}                       

\begin{document}

   \title{Gamma-ray luminosity function of Gamma-ray bright AGNs
}

   \volnopage{Vol.0 (200x) No.0, 000--000}      
   \setcounter{page}{1}          

   \author{Debbijoy Bhattacharya
      \inst{1,2}
   \and P. Sreekumar
      \inst{2}
   \and R. Mukherjee
      \inst{3}
   }

   \institute{ Department of Physics, Indian Institute of Science, Bangalore, India 560012 {\it debbijoy@gmail.com}\\
       \and
             Space Astronomy \& Instrumentation Division, ISRO Satellite Centre\\,
Bangalore, India 560017
        \and
             Barnard College, Columbia University, USA\\
   }

   \date{Received~~2009 month day; accepted~~2009~~month day}

\abstract{ Detection of $\gamma$-ray emissions from a class of active galactic nuclei (viz blazars), has been one of the important findings from the Compton Gamma Ray Observatory (CGRO). However, their $\gamma$-ray luminosity function has not been well determined. Few attempts have been made in earlier works, where BL Lacs and Flat Spectrum Radio Quasars (FSRQs) have been considered as a single source class. In this paper we investigated the evolution and  $\gamma$-ray luminosity function of FSRQs and BL Lacs separately. Our investigation indicates no evolution for BL Lacs, however FSRQs show significant evolution. Pure luminosity evolution is assumed for FSRQs and exponential and power law evolution models are examined. Due to the small number of sources, the low luminosity end index of the luminosity function for FSRQs is constrained with an upper limit. BL Lac luminosity function shows no signature of break. As a consistency check, the model source distributions derived from these luminosity functions show no significant departure from the observed source distributions. 
}

   \authorrunning{Bhattacharya, Sreekumar \& Mukherjee }            
   \titlerunning{$\gamma$-ray luminosity function of FSRQs \& BL Lacs }  

   \maketitle

%
%
\section{Introduction}           
\label{sect:intro}

Gamma-ray astronomical studies received a substantial boost after the launch of the Compton Gamma-Ray Observatory (CGRO) in 1991 (Kanbach et al.~\cite{kanbach}). Though the recently launched $\gamma$-ray missions AGILE \& FERMI Gamma-ray Space Telescope (FGST) are expected to dramatically increase the number of $\gamma$-ray sources and identify new source classes, at present the 3rd EGRET (3EG) point source catalog (Hartman et al.~\cite{b12}) provides the most complete list of GeV $\gamma$-ray sources. It contains 271 sources ($>$100 MeV), which include five pulsars, one probable radio galaxy (Cen A), 66 high confidence identifications of a sub-class of active galactic nuclei called blazars and one external normal galaxy, the Large Magellanic Cloud (LMC). In addition, 27 lower confidence potential blazar identifications are listed. Applying a different approach, Mattox, Hartman \& Reimer~(\cite{matox}) found 46 high confidence blazars (45 of these are present in the 3EG catalog) and 37 plausible candidates.   
Sowards-Emmerd, Romani \& Michelson~(\cite{sowards2003}) and Sowards-Emmerd et al.~(\cite{sowards2004}) introduced a new technique to identify $\gamma$-ray sources. Unlike earlier work (Hartman et al.~\cite{b12}; Mattox, Hartman \& Reimer~\cite{matox}), where selection has largely proceeded by correlation with an existing radio survey, Sowards-Emmerd, Romani \& Michelson~(\cite{sowards2003}) and Sowards-Emmerd et al.~(\cite{sowards2004}) attempted to obtain a more complete census of plausible blazar counterparts, sifting sources with extant radio survey data and then conducting a multiwavelength follow-up. Their technique ensured that even the plausible candidates have more than 80$\%$ good identifications. They reported 113 blazar IDs for the 3EG sources.
\par While studying any source population, it is very important to investigate their density and luminosity distribution and also their time evolution. Luminosity function of a source class gives a quantitative picture of the luminosity and density distribution of that source class. Blazars form the largest source class of identified EGRET sources, hence several papers have discussed the expected contribution of blazar emission to the diffuse $\gamma$-ray background.
\par In order to calculate their contribution to the $\gamma$-ray background, one needs to derive the source luminosity function. Luminosity function ($\phi(L,z)$) is defined as the number of sources, per unit luminosity bin ($dL$), per unit of comoving volume ($dV$),
\begin{equation}
\phi(L,z) = \frac{dN}{dVdL}
\end{equation} 
Here, we present a study of the $\gamma$-ray luminosity function of EGRET detected blazars and their evolution.

\section{Luminosity Function Construction}
\begin{figure}[h]
\centering
 \includegraphics[width=\textwidth, angle=0]{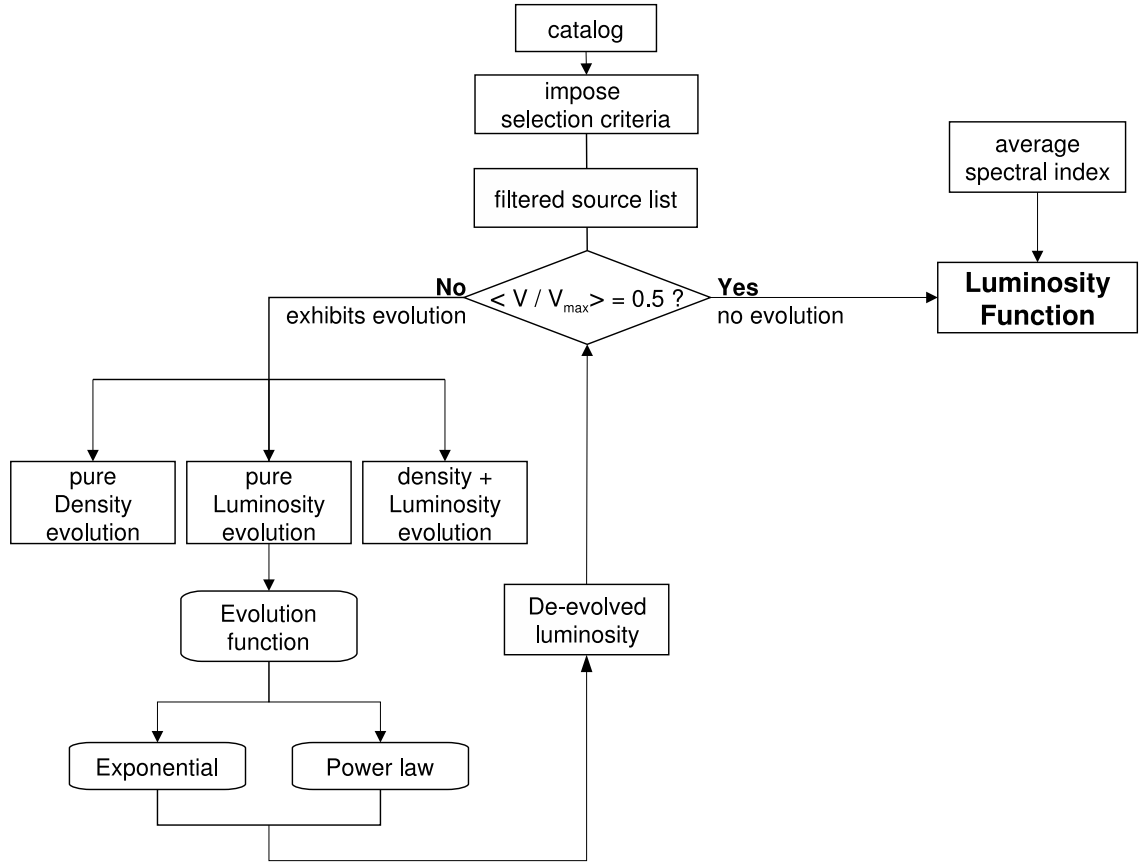} 
	\caption{Flowchart of Luminosity Function Construction} 
	\label{fig-flowchart}
\end{figure}

Luminosity function is constructed by binning sources in the luminosity and redshift plane from a complete source catalog, devoid of any selection bias. However, when the source catalog contains only a limited number of sources, the luminosity function can be constructed as follows. If the luminosity of a source in one waveband is linearly related to the luminosity in some other wavebands, one can replace the luminosity function of a given source class in one waveband by a scaled luminosity function from another waveband. This approach has been adopted in some earlier works, including Stecker, Salamon $\&$ Malkan~(\cite{b32}) and Stecker $\&$ Salamon~(\cite{b31}), who considered linear correlations between radio luminosity and $\gamma$-ray luminosity of blazars. They approximated the $\gamma$-ray luminosity functions by the radio luminosity function. A similar approach has been taken by Narumoto $\&$ Totani~(\cite{narumoto1};\cite{narumoto2}) who derived the $\gamma$-ray luminosity function from scaling the X-ray luminosity function. Alternately, Chiang et al.~(\cite{b6}) and Chiang $\&$ Mukherjee~(\cite{b5}) tried to construct the luminosity function directly from the $\gamma$-ray source catalog, without considering any correlation between radio \& $\gamma$-ray luminosity of blazars. Here we adopt a similar approach to find the luminosity functions of Flat Spectrum Radio Quasars (FSRQs) and BL Lacs separately from the latest EGRET detected blazars list. We consider a $\Lambda CDM$ cosmology (${\Omega}_{M}$=0.3, ${\Omega}_{\Lambda}$=0.7). The value of Hubble constant ($H_0$) is considered to be 70 km s$^{-1}$Mpc$^{-1}$. Fig. \ref{fig-flowchart} shows the overall scheme we followed in order to derive the luminosity function, as elaborated in the following sections.

\section{Gamma-ray Blazar Catalog}
Chiang $\&$ Mukherjee~(\cite{b5}) considered the subset of 3EG catalog sources that have a minimum radio flux of 1\,Jy $@$ 5\,GHz. There were 34 blazars in their list. After the publication of the 3EG catalog, Sowards-Emmerd, Romani \& Michelson~(\cite{sowards2003}) and Sowards-Emmerd et al.~(\cite{sowards2004}) reported new identifications of EGRET sources. They calculated the over density of sources near high galactic latitudes ($|b|\geq$ 20$^{\circ}$), which showed a considerable number of sources when the minimum radio flux is decreased from 1\,Jy to 100\,mJy. We used the list of blazar sources from Sowards-Emmerd, Romani \& Michelson~(\cite{sowards2003}) and Sowards-Emmerd et al.~(\cite{sowards2004}) catalog and separated them into BL Lacs and FSRQs. While constructing a gamma-ray blazar sample, the cutoff in the source detection significance limit has been taken as  4$\sigma$ for sources above the Galactic plane $(|b| > 10^{\circ})$ and as 5$\sigma$ for sources in the Galactic plane $(-10^{\circ} \leq b \leq 10^{\circ})$. The final list includes 46 FSRQs and 15 BL Lacs in our sample. Three BL Lacs do not have redshift information. These three sources are only used to calculate the average spectral index and normalization of the luminosity function. 
\par We adopted the $\frac{V}{V_{max}}$ test (Avni \& Bahcall~\cite{avni}) in order to test for any source evolution. Here $V$ is the volume enclosed at the known distance of the source. There exists a maximum ($z_{max}$) distance at which the source is at the limiting flux of the survey. $V_{max}$ is the volume corresponding to this maximum distance. Beyond $z_{max}$ the source flux falls below the flux limit of the survey and hence cannot be detected. For a system with no evolution, $\frac{V}{V_{max}}$ values should be uniformly distributed between 0 and 1 and hence, $<\frac{V}{V_{max}}>$ = 0.5. Since the sample is flux limited both in radio and $\gamma$-rays, each source is assigned a $z_{max}$ = min($z_{max,radio},z_{max,{\gamma}}$). While calculating $z_{max,radio}$ for FSRQs, the radio evolution function of Dunlop and Peacock~(\cite{dunlop}) is used ( $f_{DP}(z)  =  10^{(az + bz^2)}$, where $a$ = 1.18 \&  $b$ = -0.28 ).
For BL Lacs, the radio evolution function from Stickel et al.~(\cite{stickel}) is used ( $ exp[\frac{T(z)}{{\tau}_R}]$, where $T(z)$ is the look-back time and ${\tau}_R$ is the evolutionary time scale in the units of Hubble time ). 
\par Our sample contains 12 BL Lac sources. For BL Lacs, the value of  $<\frac{V}{V_{max}}>$ is 0.59 $\pm$0.08. Error in $<\frac{V}{V_{max}}>$ is estimated using $\sigma = (12 N)^{-\frac{1}{2}}$, where $N$ is the number of sources in the sample.
\begin{figure}[h!]
\begin{minipage}[t]{0.495\linewidth}
  \centering
   \includegraphics[width=59mm,height=58mm]{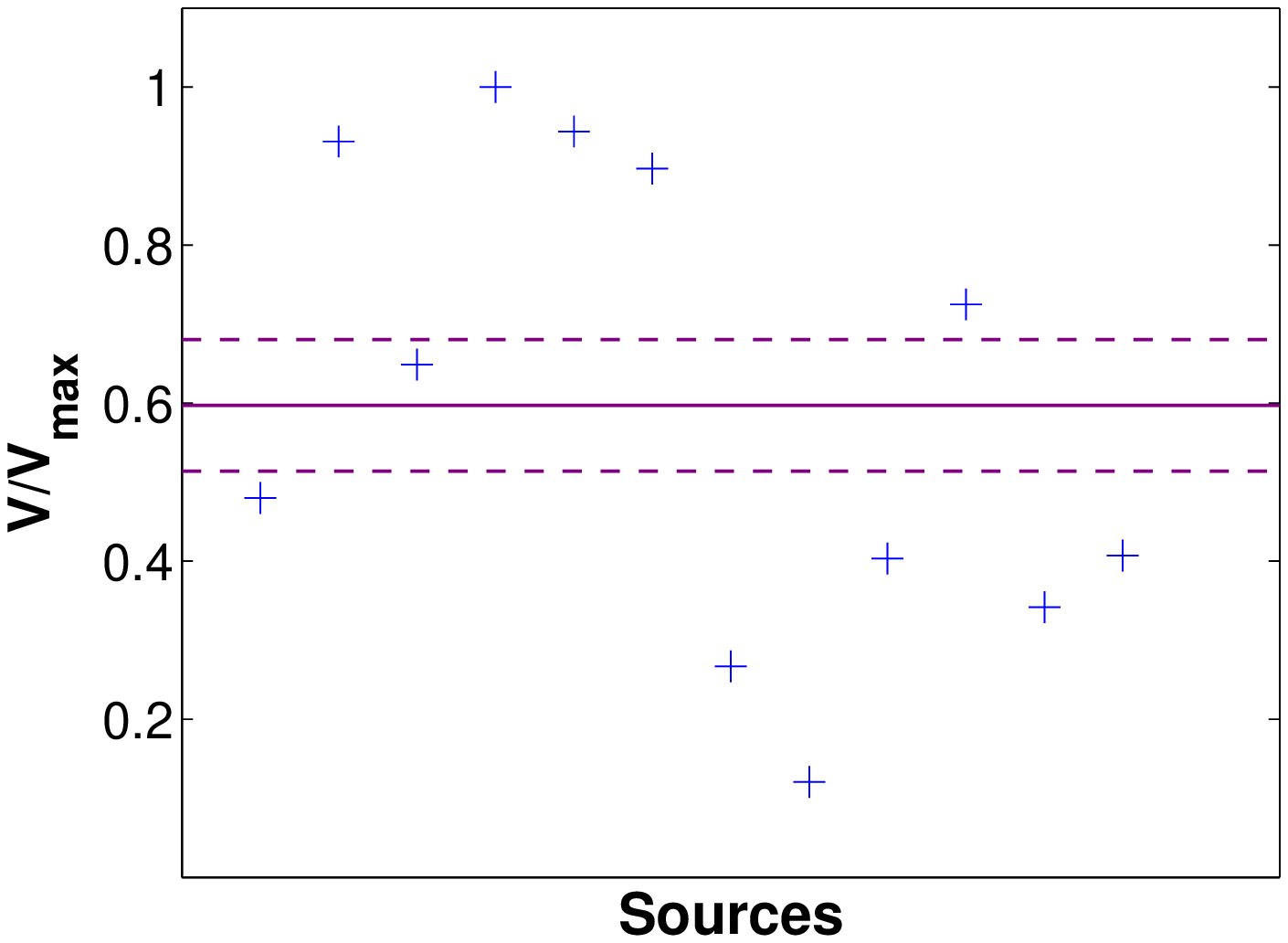} 


        	\caption{Distribution of $\frac{V}{V_{max}}$ for BL Lacs} 
\label{fig-bllacvvm1}
\end{minipage}%
  \begin{minipage}[t]{0.495\textwidth}
  \centering
   \includegraphics[width=59mm,height=58mm]{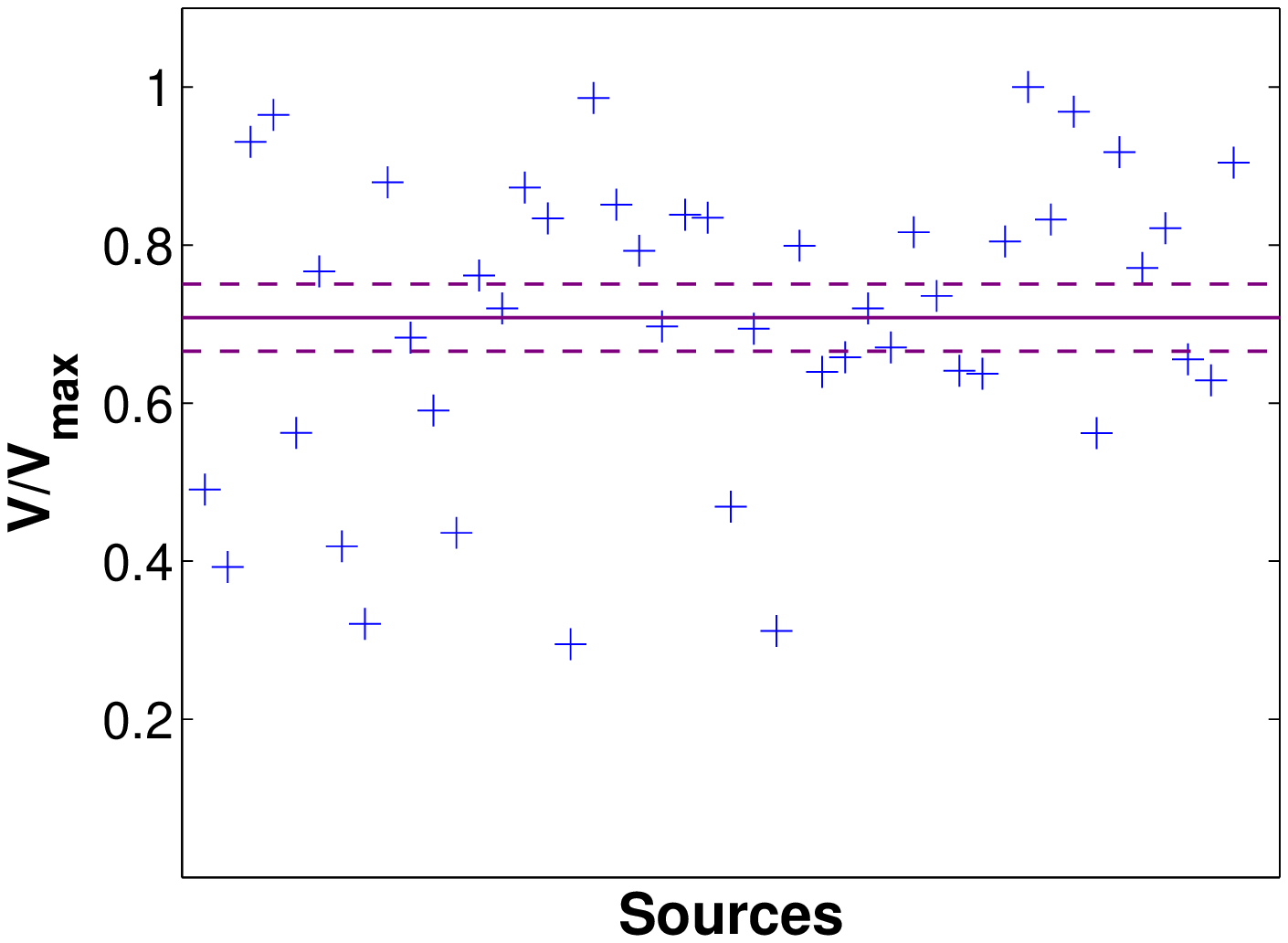}
\caption{Distribution of $\frac{V}{V_{max}}$ for FSRQs}
\label{fig-fsrqvvmnoevo}
\end{minipage}%

\end{figure}
Hence we conclude from the $\frac{V}{V_{max}}$ test that BL Lacs show no measurable evolution. We also compare the $\frac{V}{V_{max}}$ distribution with a uniform distribution by KS-test and Quantile-Quantile plot, which also shows that the distribution is uniform. There are 46 FSRQs in our sample. For FSRQs, the value of $<\frac{V}{V_{max}}>$ is $0.71 \pm  0.04 $, which indicates strong evolution (5.2$\sigma$). The $\frac{V}{V_{max}}$ distribution of BL Lacs and FSRQs are shown in Fig. \ref{fig-bllacvvm1} and Fig. \ref{fig-fsrqvvmnoevo} respectively.
\par Any source class can exhibit pure luminosity evolution where only the source luminosity changes with redshift, while the source density remains constant. Alternately, it can exhibit pure density evolution where only the source density varies with redshift. More realistically, both luminosity and density evolution can be expected. Considering the limited number of $\gamma$-ray sources, it is difficult to examine an evolution model that incorporates both luminosity \& density evolution. We examined two pure density evolution models, $[ (1+z)^{\beta_1}$ and $exp(\beta_2 T(z)) ]$. We found large errors in the density parameter values as derived from our source list ($\beta_1 = 5.8^{+ 1.6}_{-1.5}$ and $\beta_2 = 10.9^{+ 3.0}_{-2.8}$). These large errors prevents us from drawing useful conclusions. Since pure luminosity evolution is more often observed at other wavelengths, we examine such a model for $\gamma$-ray blazars.  

\section{Luminosity Evolution Function for FSRQs}
$\frac{V}{V_{max}}$ analysis shows a clear indication of evolution of FSRQs in $\gamma$-rays. We consider the pure luminosity evolution  (density of sources is constant with $z$) of these sources. The luminosity of a source at a redshift $z$ can be written as 
\begin{equation}
L(z) = L_0\times f(z)
\end{equation}
where $L_0$ is the luminosity at zero redshift and $f(z)$ is the luminosity evolution function.
\par
\begin{figure}[h]
\begin{minipage}[t]{0.495\linewidth}
	\centering
\includegraphics[width=59mm,height=58mm]{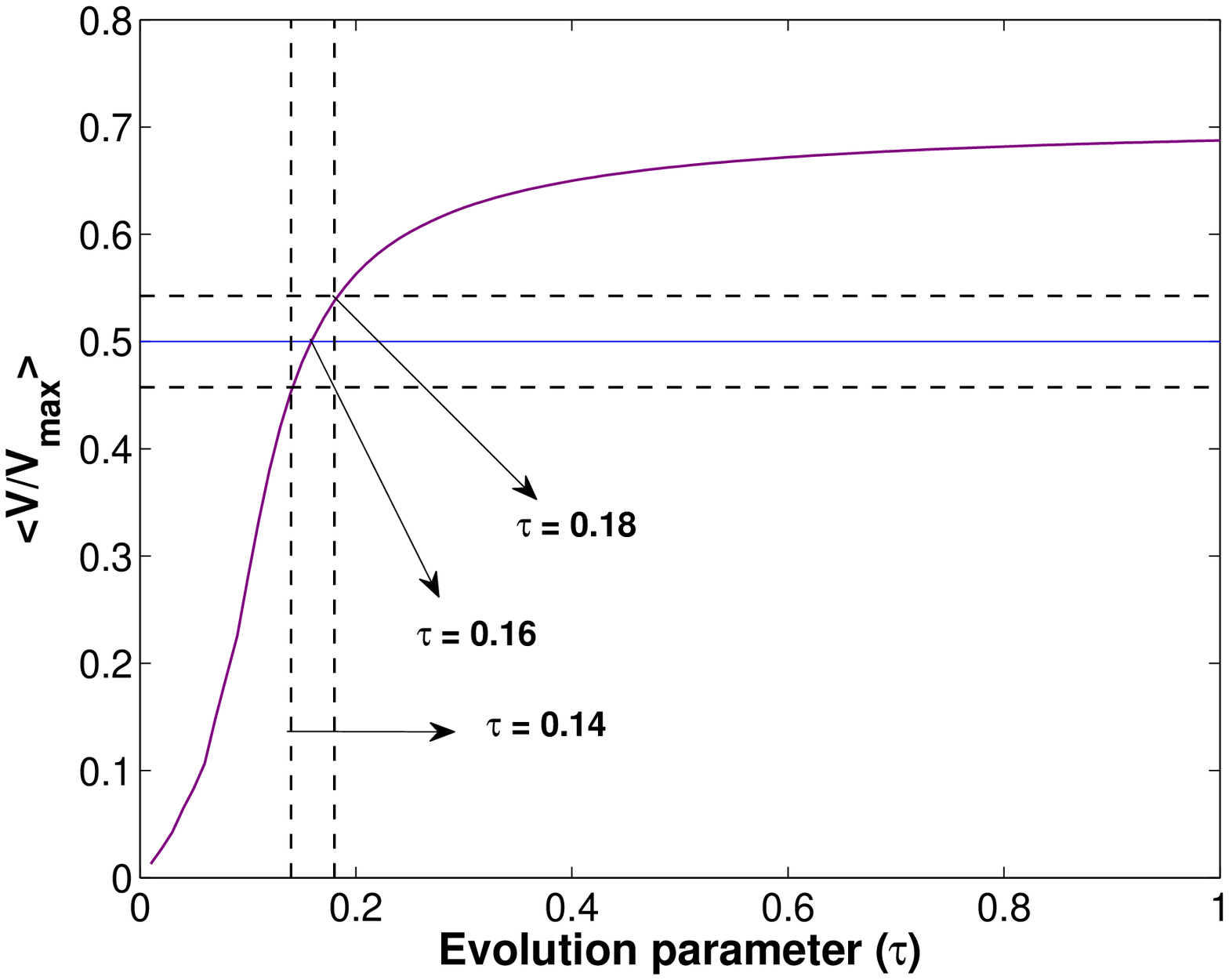} 
	\caption{{\small Distribution of $<\frac{V}{V_{max}}>$ with different $\tau$ values. Dashed horizontal lines indicate the 1 $\sigma$ error in $<\frac{V}{V_{max}}>$. Dashed vertical lines show the range of $\tau$ values for which $\frac{V}{V_{max}}$ distribution becomes uniform within 1 $\sigma$ error. }} 
	\label{fig-tau-vvm}
\end{minipage}%
  \begin{minipage}[t]{0.495\textwidth}
\centering
\includegraphics[width=59mm,height=58mm]{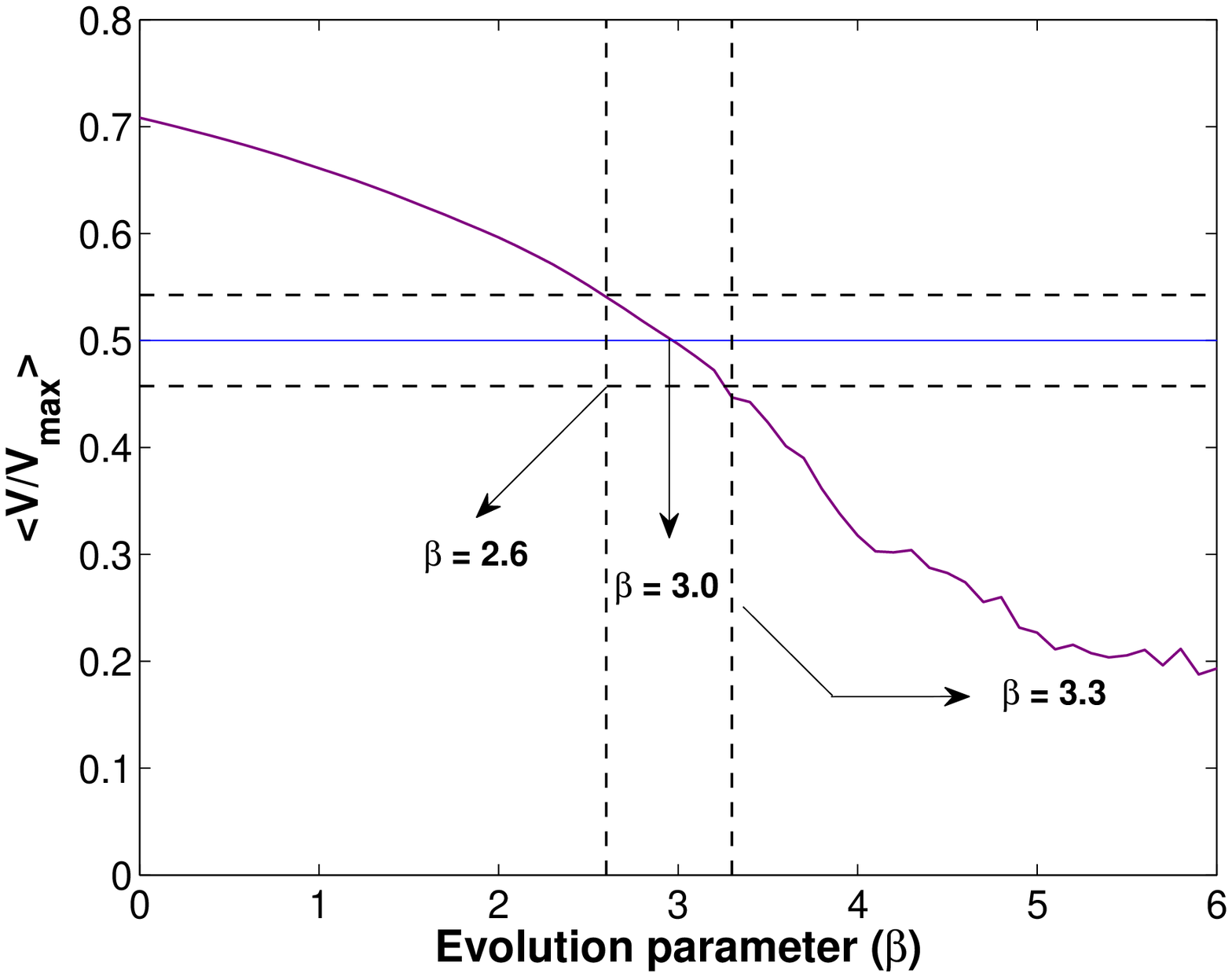} 
\caption{{\small Distribution of $<\frac{V}{V_{max}}>$ for FSRQs with different $\beta$ values. Dashed horizontal lines indicate the 1 $\sigma$ error in $<\frac{V}{V_{max}}>$. Dashed vertical lines show the range of $\beta$ values for which $\frac{V}{V_{max}}$ distribution becomes uniform within 1 $\sigma$ error.}} 
\label{fig-betavvmf46}
\end{minipage}
\end{figure}

 Two types of luminosity evolution functions, exp($\frac{T(z)}{\tau}$) and $(1+z)^{\beta}$ are considered. Here $T(z)$ denotes look-back time. We used the modified $\frac{V}{V_{max}}$ method to find the evolution parameters ${\tau}$ and $\beta$. For the optimum parameter value, $\frac{V}{V_{max}}$ should be uniformly distributed between 0 and 1, thus $<\frac{V}{V_{max}}> = 0.5$. For very high values of the evolution parameter, we sometimes get $z_{min}$ as well as $z_{max}$. The physical significance of $z_{min}$ is that, for that particular evolution parameter value, the source cannot be observed below $z_{min}$. For all cases with $z_{max}$ $\ge$ $5$ we assumed $z_{max}$ $= 5$. Fig. \ref{fig-tau-vvm} shows the variation of $<\frac{V}{V_{max}}>$ with different $\tau$ values. Horizontal dashed lines show the 1 $\sigma$ error in $<\frac{V}{V_{max}}>$. For each value of the evolution parameter ($\tau$), the distribution of $\frac{V}{V_{max}}$ is compared with a uniform distribution. We find $\tau = 0.16 \pm 0.02$. KS-test is performed which shows the distribution of $\frac{V}{V_{max}}$ is uniform for $\tau = 0.16 \pm 0.02$. We randomly took 46 points from a uniform distribution and studied the Quantile-Quantile plot of the $\frac{V}{V_{max}}$ (after de-evolution for $\tau$ = 0.16) with these 46 randomly chosen points. It shows no significant departure from linearity and thus again demonstrates that the $\frac{V}{V_{max}}$ distribution is uniform.
Fig. \ref{fig-betavvmf46} shows the variation of  $<\frac{V}{V_{max}}>$ with $\beta$ considering the second evolutionary model (power law). We find $\beta = 3.0^{+ 0.3}_{-0.4}$ for which $<\frac{V}{V_{max}}>$ becomes 0.50 (within 1 $\sigma$). 
KS-test \& Quantile-Quantile plot also show the distribution is uniform for $\beta = 3.0^{+ 0.3}_{-0.4}$.

\par For both models of evolution, all luminosities have been de-evolved to $z = 0$ and these are used to construct the de-evolved luminosity function by the standard nonparametric $\frac{1}{V_{max}}$ method and the likelihood method. We used the average spectral index of the sources for luminosity function determination. 

\section{Spectral index distribution of FSRQs and BL Lacs}
\begin{figure}[h]
\begin{minipage}[t]{0.495\textwidth}
	\centering
\includegraphics[width=59mm,height=58mm]{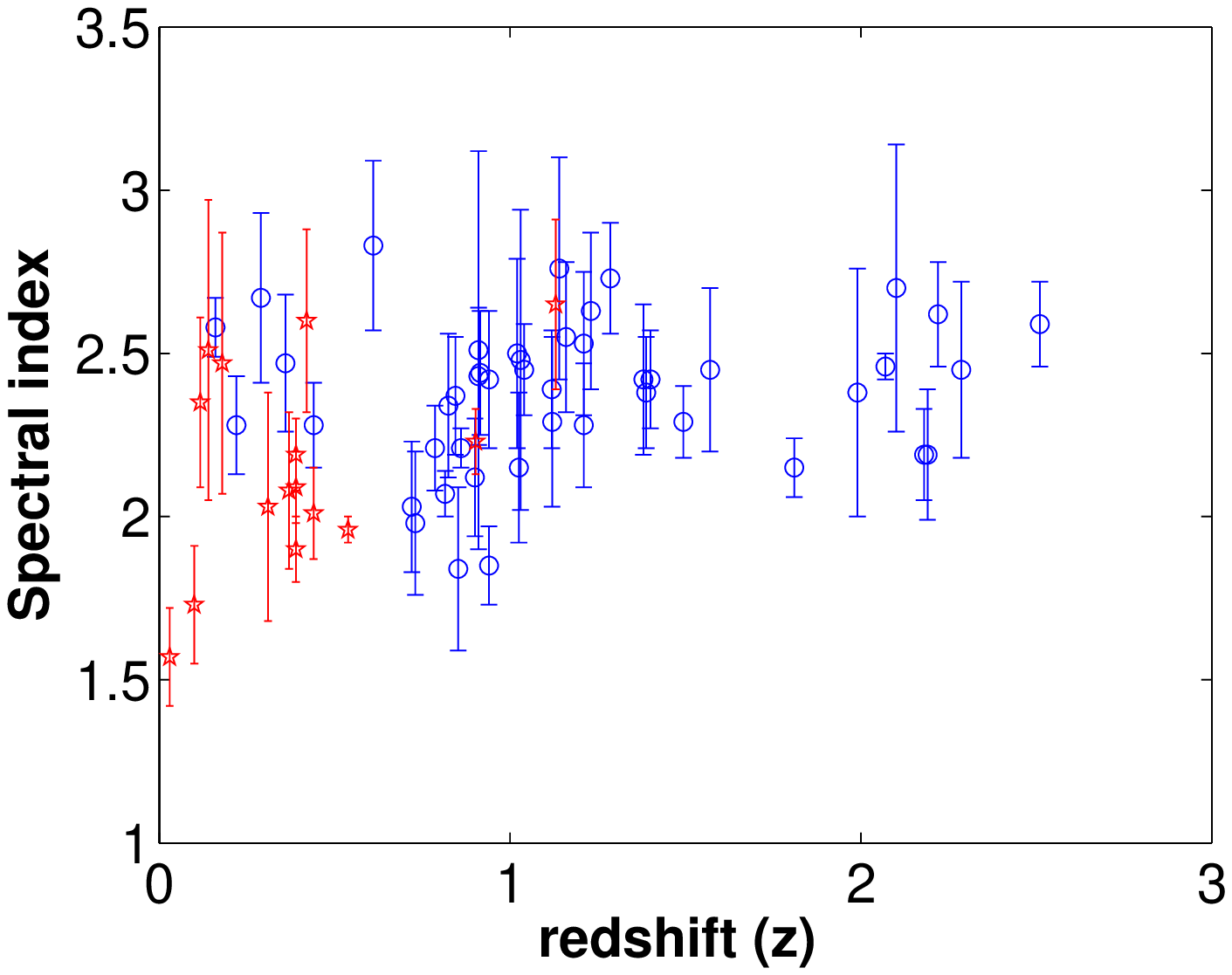}
	\caption{{\small Spectral index distribution with redshift. The circles are the FSRQ objects and the diamonds are the BL Lac objects.}} 
	\label{fig-spz}
\end{minipage}%
  \begin{minipage}[t]{0.495\textwidth}
	\centering
\includegraphics[width=59mm,height=58mm]{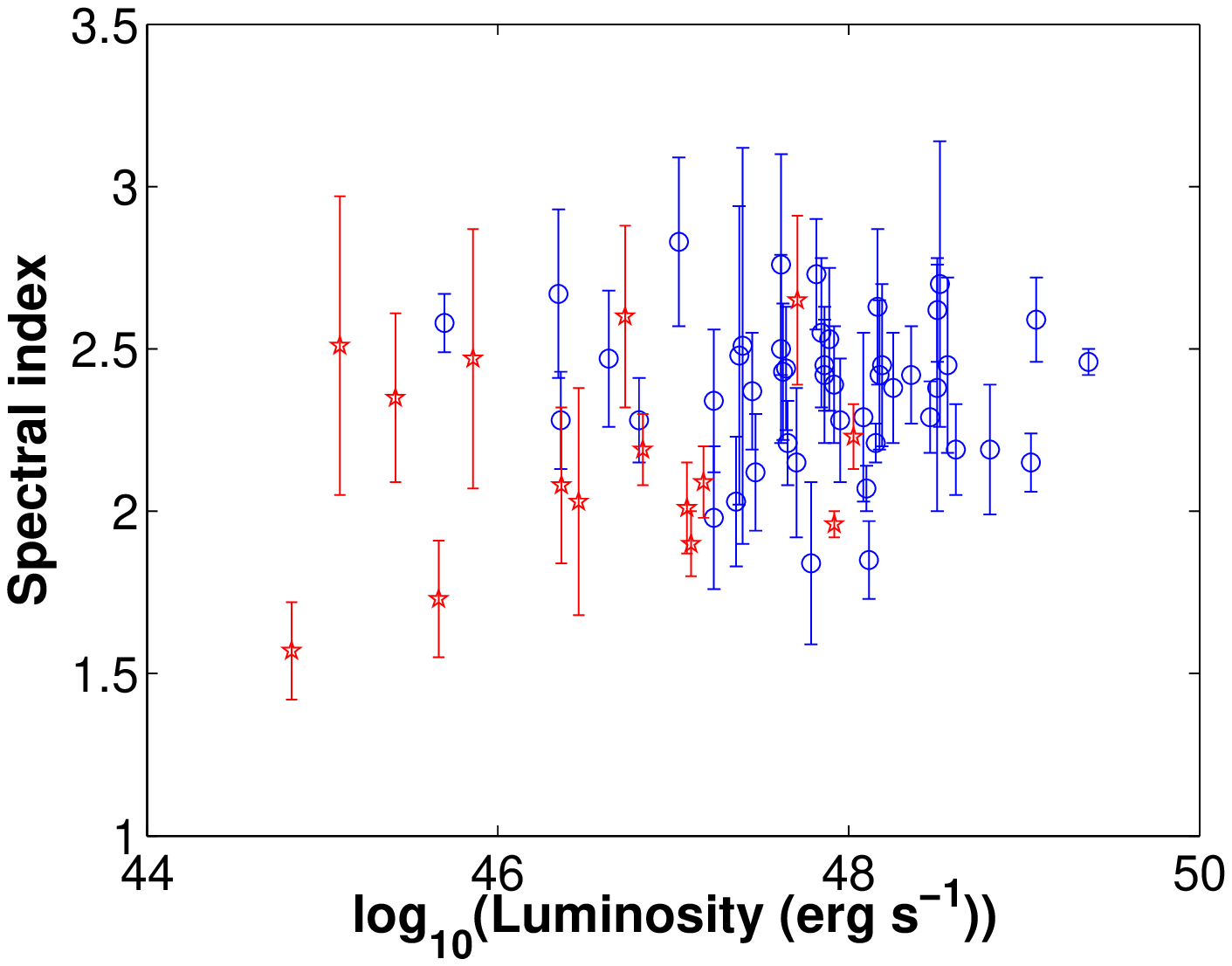}
	\caption{{\small Spectral index distribution with luminosity. The circles are the FSRQ objects and the diamonds are the BL Lac objects.}} 
	\label{fig-spdelumnoevo}
\end{minipage}
\end{figure}

The photon spectral index distribution with redshift of FSRQs and BL Lacs from our source list are shown in Fig. \ref{fig-spz}. The photon spectral index distribution of these objects with luminosity are shown in Fig. \ref{fig-spdelumnoevo}. In Fig. \ref{fig-spdelumnoevo}, no evolution of FSRQs and BL Lacs have been considered. Of the 15 BL Lac sources, the three BL Lac sources without redshift information are considered to have the average redshift of the distribution.
To derive the average $\gamma$-ray spectral index, we assume that the intrinsic spectral index distribution (ISID) can be described by a Gaussian (Venters $\&$ Pavlidou~(\cite{venters2007}),
\begin{equation}
\mbox{ISID}(\alpha)d\alpha = \frac{1}{\sqrt{2\pi}\sigma_0}exp\bigg[- \frac{(\alpha - \alpha_0)^2}{2\sigma_{0}^{2}}\bigg]d\alpha
\end{equation}
The likelihood function of spectral index distribution has been adopted from Venters $\&$ Pavlidou~(\cite{venters2007}).
 \begin{figure}[h]
\begin{minipage}[t]{0.495\linewidth}
	\centering
\includegraphics[width=59mm,height=58mm]{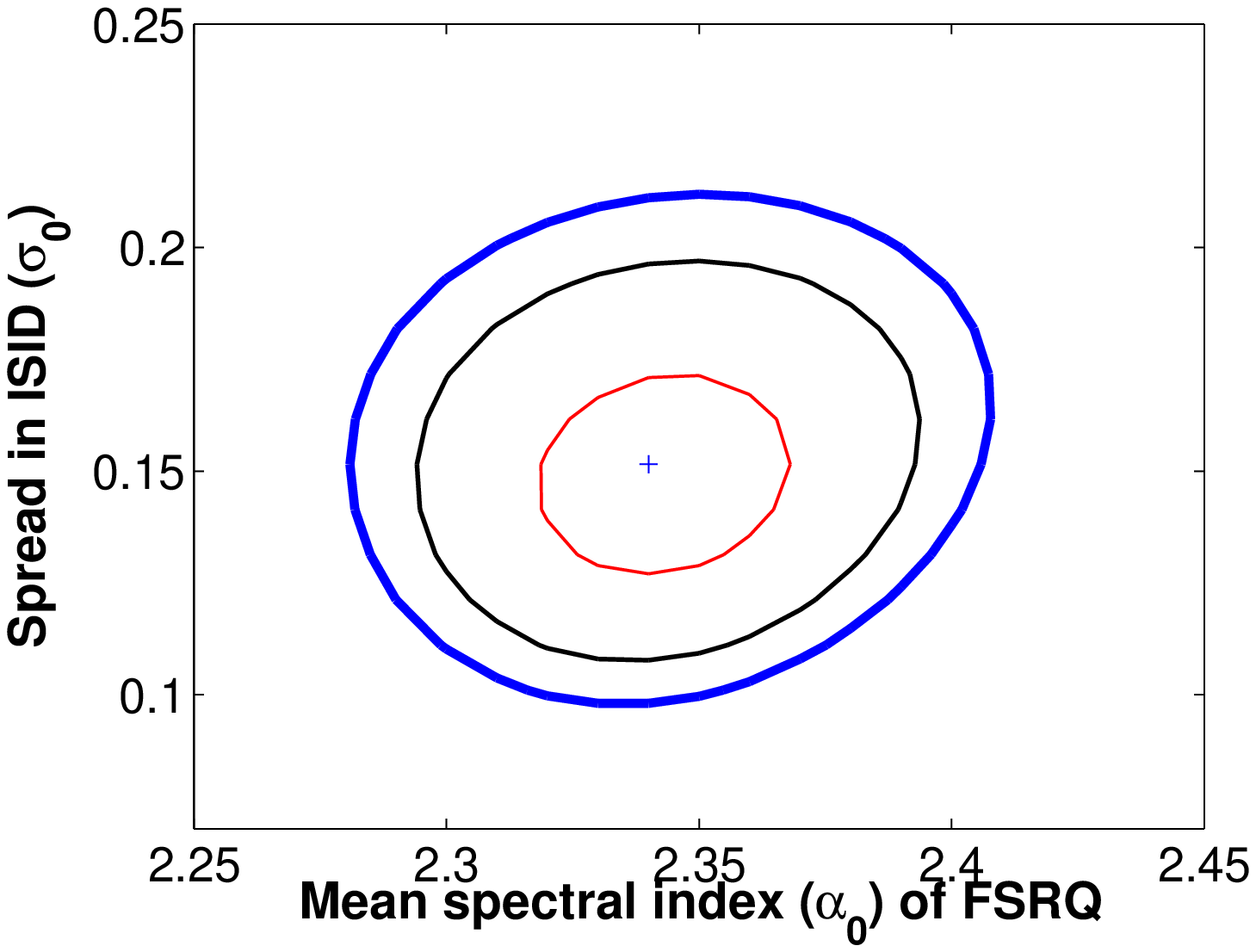}
  	\caption{{\small $68\%$, $95\%$ and $99\%$ likelihood function contours for FSRQs}} 
	\label{fig-spinfsrqc}
\end{minipage}
\begin{minipage}[t]{0.495\textwidth}
	\centering
\includegraphics[width=59mm,height=58mm]{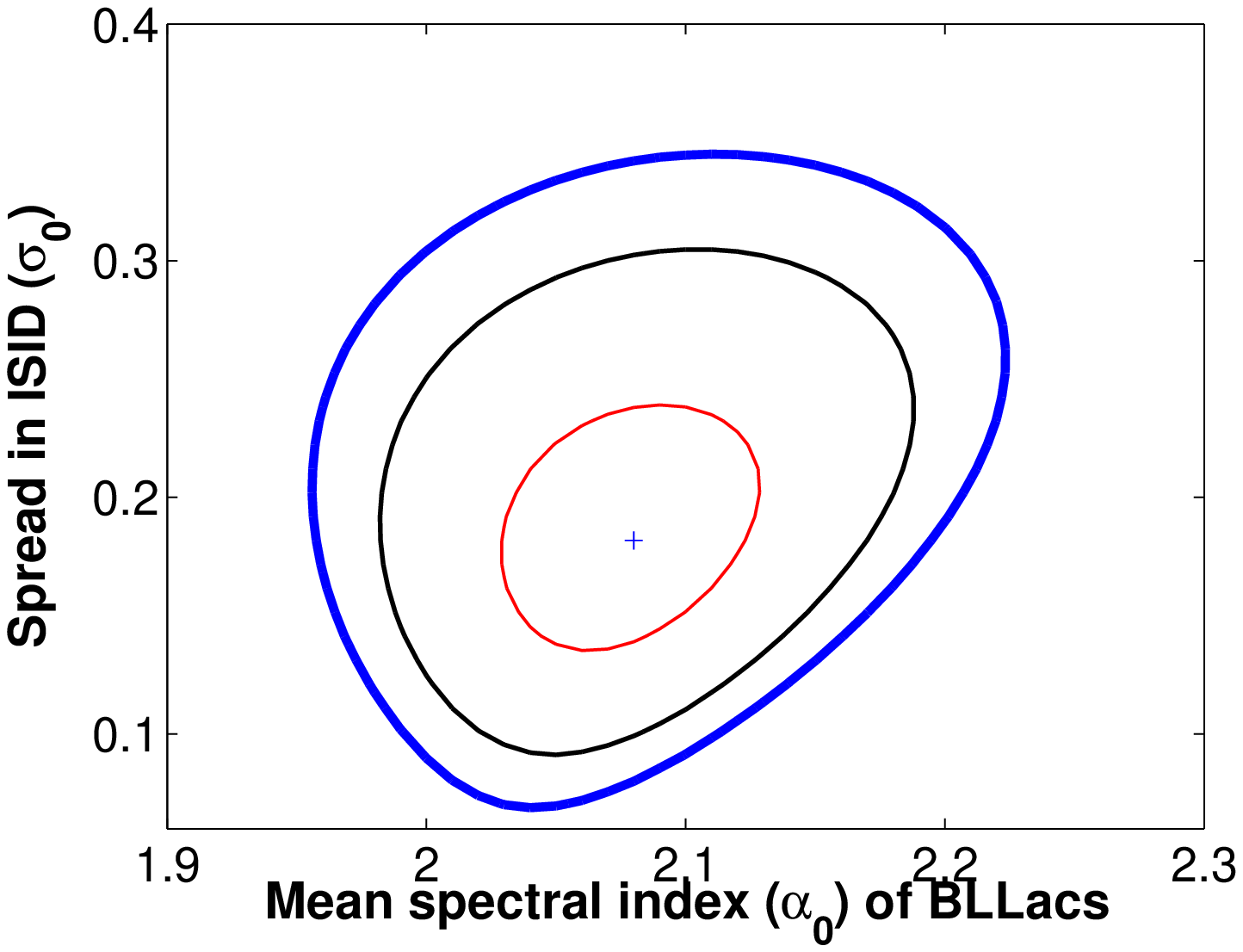}
	\caption{{\small $68\%$, $95\%$ and $99\%$ likelihood function contours for BL Lacs}} 
	\label{fig-spinbllac1c}
\end{minipage}
\end{figure}

Fig. \ref{fig-spinfsrqc} and Fig. \ref{fig-spinbllac1c} show the $68\%$, $95\%$ and $99\%$ likelihood function contours of ${\alpha}_0$ and ${\sigma}_0$ for FSRQs and BL Lacs respectively. Venters $\&$ Pavlidou (2007) also calculated the average spectral index for FSRQs and BL Lacs. However, their source selection criteria did not require $> 4 \sigma$ detection of sources when averaged over the first four phases of EGRET observations. The values of ${\alpha}_0$ \& ${\sigma}_0$ of our calculations and from Venters $\&$ Pavlidou~(\cite{venters2007}), are given in Table \ref{Tab:publ-spindexblazar}.
\begin{table}
\begin{center}
\caption{The Average $\gamma$-ray Spectral Index Value for Blazars}\label{Tab:publ-spindexblazar}


 \begin{tabular}{clcl}
  \hline\noalign{\smallskip}
 & This work & Venters \& Pavlidou (2007)   \\
  \hline\noalign{\smallskip}
FSRQ  & 2.34 $\pm$ 0.15 & 2.30 $\pm$ 0.19  \\ 
BL Lac  & 2.08 $\pm$ 0.18 & 2.15 $\pm$ 0.28  \\
\noalign{\smallskip}\hline
\end{tabular}
\end{center}
\end{table}

\section{Luminosity Function Construction of FSRQs}
After de-evolution of the source luminosities at $z = 0$, we constructed their luminosity function by the following methods.
\newline
(a) The luminosity function can be written as
\begin{equation}
\phi(L_0) = \frac{1}{dL_0}\sum^{N}_{i=1} \frac{1}{V_{max}(i)}
\end{equation}
Here, $L_0$ is the de-evolved luminosity. For our data set, $\frac{1}{V_{max}}$ method only gives the upper-end index of the luminosity function. The break luminosity (if any) and the lower end index cannot be found by this method.
\\
(b) Maximum Likelihood function has been generated for the distribution of sources, considering a broken power law luminosity function. The high luminosity end slope has been taken from the non-parametric $\frac{1}{V_{max}}$ method (as discussed in (a) ). The break luminosity and the lower end index are the free parameters, and they are constrained using the likelihood function.

\subsection{$\frac{1}{V_{max}}$ Method}
\begin{figure}[h]
\begin{minipage}[t]{0.495\linewidth}
	\centering
\includegraphics[width=59mm,height=58mm]{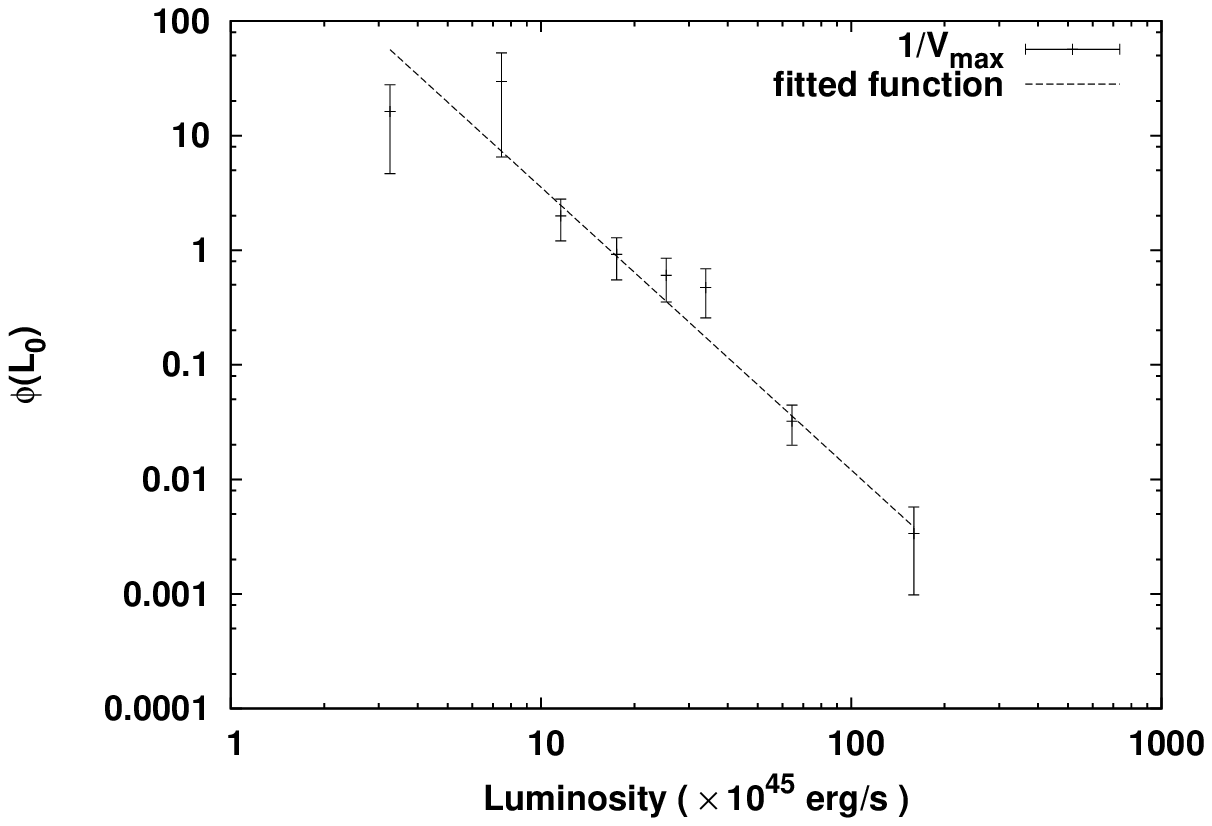}
	\caption{{\small De-evolved luminosity function ($\phi(L_0)$) of FSRQs ($\tau$ = 0.16) ($\frac{1}{V_{max}}$ method). $\phi(L_0)$ is given in $(\frac{3}{4\pi})\times (\frac{H_0}{c})^{3}$ per luminosity (10$^{45}$ erg s$^{-1}$ unit) interval.}} 
	\label{fig-lvvmtau0.16}
\end{minipage}
\begin{minipage}[t]{0.495\textwidth}
	\centering
\includegraphics[width=59mm,height=58mm]{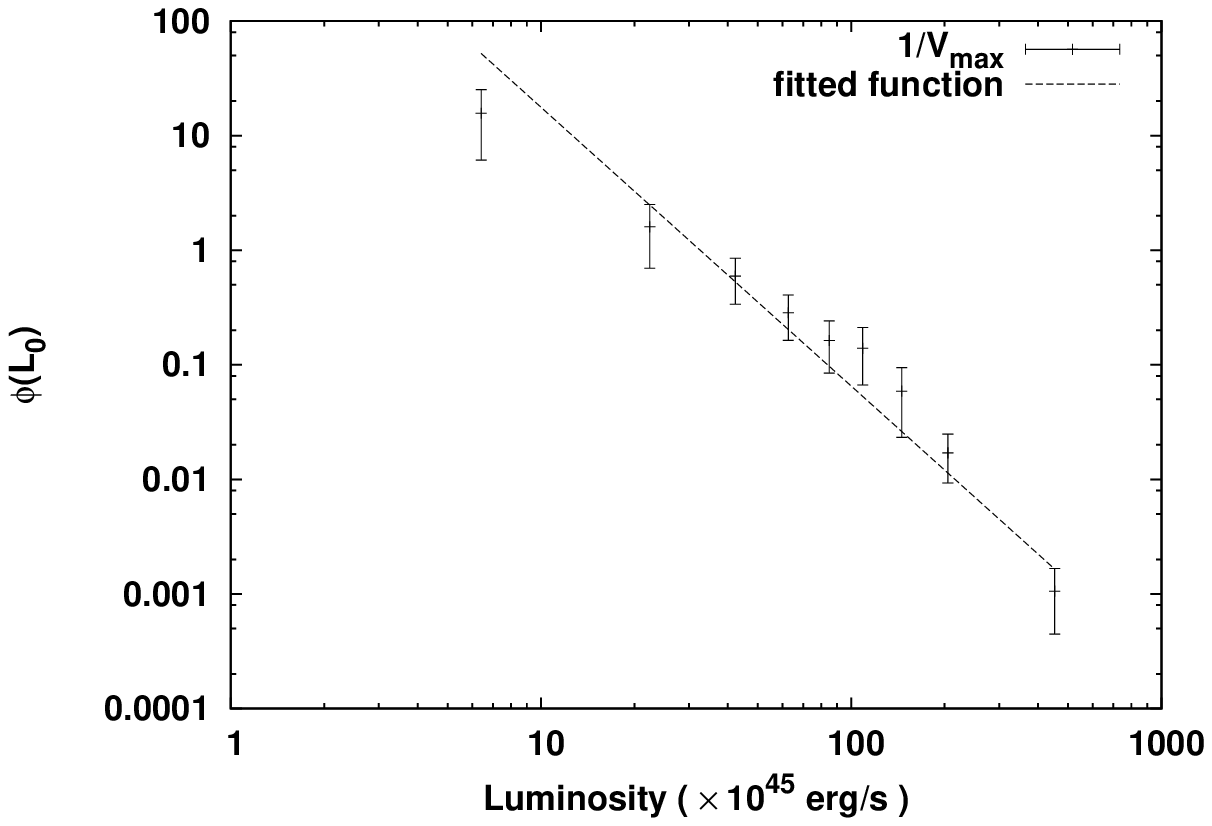}
	\caption{{\small De-evolved luminosity function ($\phi(L_0)$) of FSRQs ($\beta$ = 3.0) ($\frac{1}{V_{max}}$ method). $\phi(L_0)$ is given in $(\frac{3}{4\pi})\times (\frac{H_0}{c})^{3}$ per luminosity (10$^{45}$ erg s$^{-1}$ unit) interval.}} 
	\label{fig-lvvmbeta3.046}
\end{minipage}
\end{figure}

The non-parametric $\frac{1}{V_{max}}$ method has been employed in order to find FSRQ luminosity function ($\phi(L_0)$). FSRQs have been binned into different luminosity intervals. The high end part of the luminosity function is well fitted by a power law with index of 2.5$\pm$ 0.2 (assuming exponential evolution function with $\tau = 0.16$). For power law evolution function ($\beta = 3.0$), the best fitted value of the power law index is 2.4 $\pm$ 0.2. At the low luminosity end there is a hint of a turn over, but data points are insufficient to confirm it unambiguously.  
\newline Fig. \ref{fig-lvvmtau0.16} \& Fig. \ref{fig-lvvmbeta3.046} show the variation of $\phi(L_0)$ (determined by $\frac{1}{V_{max}}$ method) with de-evolved luminosity considering the exponential evolution function  and power law evolution function respectively.

\par In order to examine a possible change of index at the low luminosity end and to find the break luminosity (if any), we return to the source redshift distribution. Maximum likelihood analysis is used to derive the break luminosity and the low-luminosity end index.

\subsection{Maximum Likelihood Analysis}
We assumed a broken power law form of the de-evolved luminosity function.
\begin{eqnarray}
\phi(L_0) & = &  {\phi}_0 \times \Bigg (\frac{L_0}{L_B}\Bigg )^{-{\alpha}_1} \, \, \, \, L_0 \le L_B \, , \nonumber \\
& = & {\phi}_0 \times \Bigg(\frac{L_0}{L_B}\Bigg )^{-{\alpha}_2} \, \, \, \, L_0 > L_B \, .
\end{eqnarray}
Here $\phi(L_0)$ $(= \frac{dN}{dVdL_0})$ is the de-evolved luminosity function and $\phi_0$ is the normalization of the luminosity function. $L_B$ is the break luminosity. We fixed the high luminosity end index (${\alpha}_2$) of the luminosity function from $\frac{1}{V_{max}}$ method (as described in the last subsection). The redshift distribution of EGRET detected FSRQs is used to find both break luminosity ($L_B$) and the power law index ($\alpha_1$) of the low-luminosity end of the $\phi(L_0)$. The likelihood function for the redshift distribution of FSRQs (similar to Chiang \& Mukherjee~\cite{b5}) is given by
\begin{equation}
\mathcal{L}  =  \prod_{i}^{N} \frac{[dN/dz]_{i}(z_i)}{ \int_{0}^{z_{max}} [dN/dz]_{i}(z) dz} 
\end{equation} 
 Source distribution with $z$ is given by,
\begin{equation}
 \bigg[ \frac{dN}{dz} \bigg]_{i} (z) = \Theta(z) \int_{L_{i,lim}(z)}^{\infty} \frac{dV}{dz} \phi(L_0) dL_0
\end{equation}
Here $L_{i,lim}(z)$ is given by,
\begin{equation}
L_{i,lim}(z) = \frac{4 \pi F_{i,lim} D_{L}^{2}(z)}{(1 + z)^{1- \alpha_{\gamma}} f(z)}
\end{equation}

$f(z)$ is the $\gamma$-ray evolution function and ${\alpha}_{\gamma}$ is the average $\gamma$-ray energy spectral index of FSRQs. $F_{i,lim}$ is the limiting flux of the $i$th source (Chiang et al.~\cite{b6}). $\Theta(z)$ is the radio completeness function as defined in Chiang $\&$ Mukherjee~(\cite{b5}), which is the fraction of the FSRQs at a given redshift $z$ having radio flux greater than the radio limiting flux of the survey (100\,mJy).

\begin{eqnarray}
\Theta(z) & = & \int_{P_{lim,radio}(z)}^{\infty} dP \frac{dN}{dP} \bigg(\int_{0}^{\infty}  dP \frac{dN}{dP}\bigg)^{-1}\\
 \mbox{where,} & & \frac{dN}{dP} \propto \frac{1}{P} \big [ (\frac{P}{P_B})^{0.83} + (\frac{P}{P_B})^{1.96}\big ]^{-1} \quad \mbox{[Dunlop and Peacock (1990)]} \nonumber
\end{eqnarray}

Here $P_B = 10^{25.26}$\, W\,Hz$^{-1}$\,sr$^{-1}$

\begin{equation}
P_{i,lim,radio}(z)  =  \frac{F_{lim,radio} D_{L}^{2}(z)}{(1 + z)^{1 - {\alpha}_{radio}} f_{DP}(z)}
\end{equation}
   The de-evolved limiting luminosity $P_{i,lim,radio}(z)$ is a function of $z$ and varies from source to source. The radio evolution function is taken from Dunlop and Peacock~(\cite{dunlop}).
\par
\begin{figure}[h]
\begin{minipage}[t]{0.495\linewidth}
	\centering
\includegraphics[width=59mm,height=58mm]{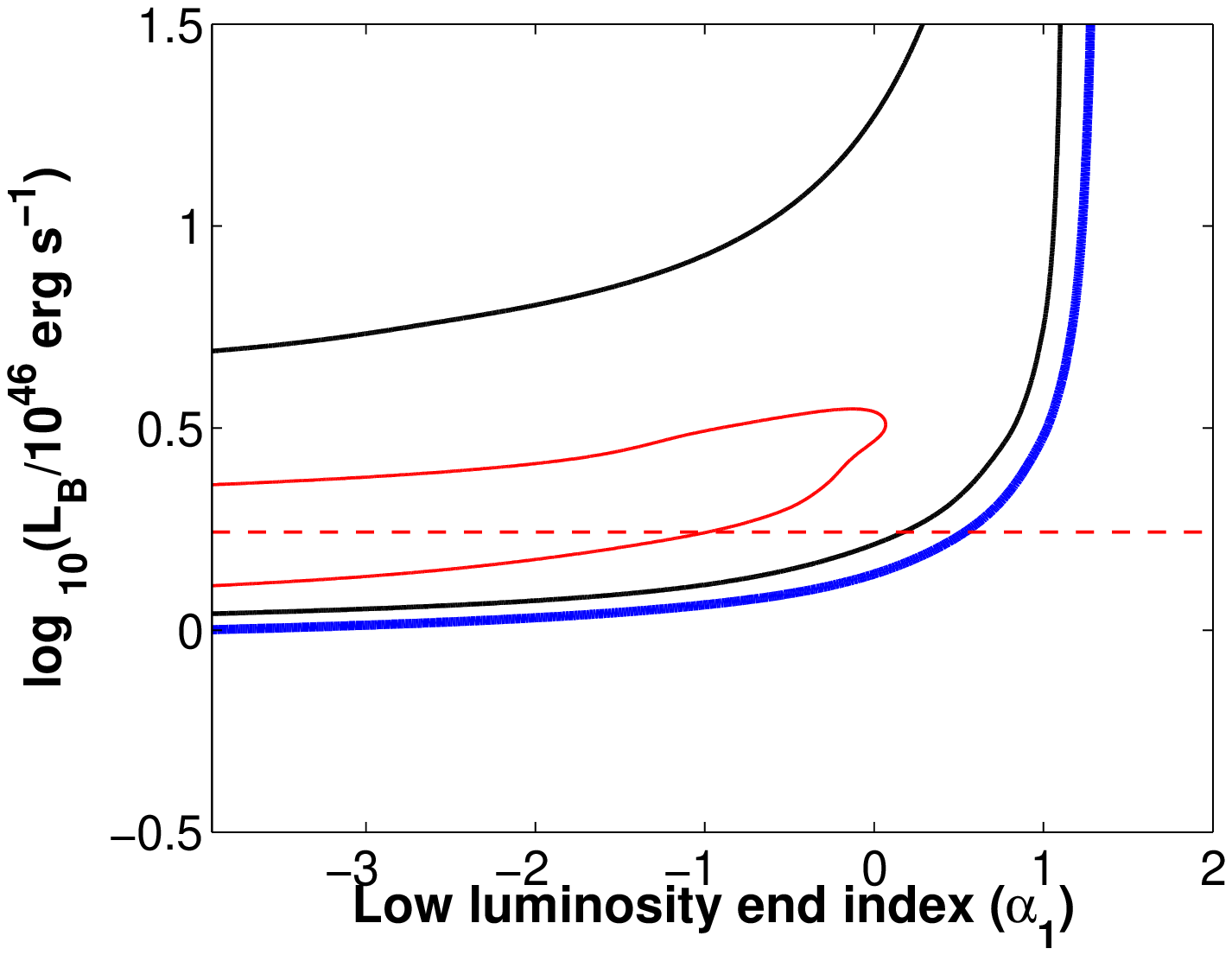}
	\caption{{\small $68\%$,$95\%$,$99\%$ contours considering exponential evolution function. Dashed line shows the maximum likelihood $L_B$ value}} 
	\label{fig-fqlbg1tau0.1646c}
\end{minipage}
\begin{minipage}[t]{0.495\textwidth}
	\centering
\includegraphics[width=59mm,height=58mm]{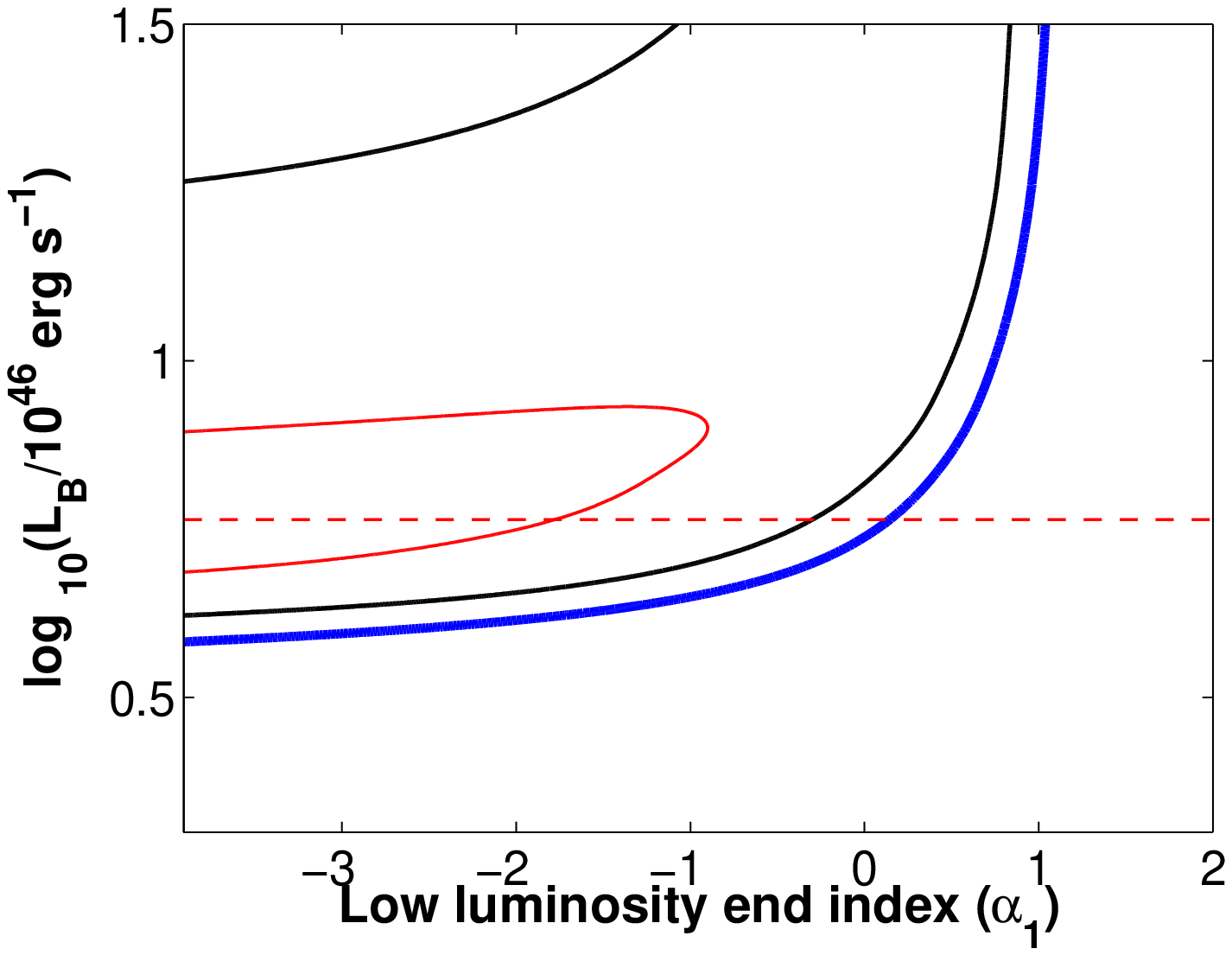}
	\caption{{\small $68\%$,$95\%$,$99\%$ contours considering power law evolution function. Dashed line shows the maximum likelihood $L_B$ value }} 
	\label{fig-fqlbg1beta3.046g2.4c}
\end{minipage}
\end{figure}

 For an exponential evolution function, we find a break luminosity ($L_B$) of $1.7 \times 10^{46}$\,erg\,s$^{-1}$.This analysis gives an upper limit to the low luminosity end power law index $\alpha_1$. Fig. \ref{fig-fqlbg1tau0.1646c} gives the $99 \%$ upper limit of $\alpha_1$ as 1.3 and the $95 \%$ upper limit as 1.1. For the power law evolution function, we find break luminosity $L_B$ is $5.8 \times 10^{46}$\,erg\,s$^{-1}$. Fig. \ref{fig-fqlbg1beta3.046g2.4c} gives the $99 \%$ upper limit of $\alpha_1$ as 1.1 and the $95 \%$ upper limit as 0.9.

\begin{figure}[h!]
	\centering
	\includegraphics[width = \textwidth]{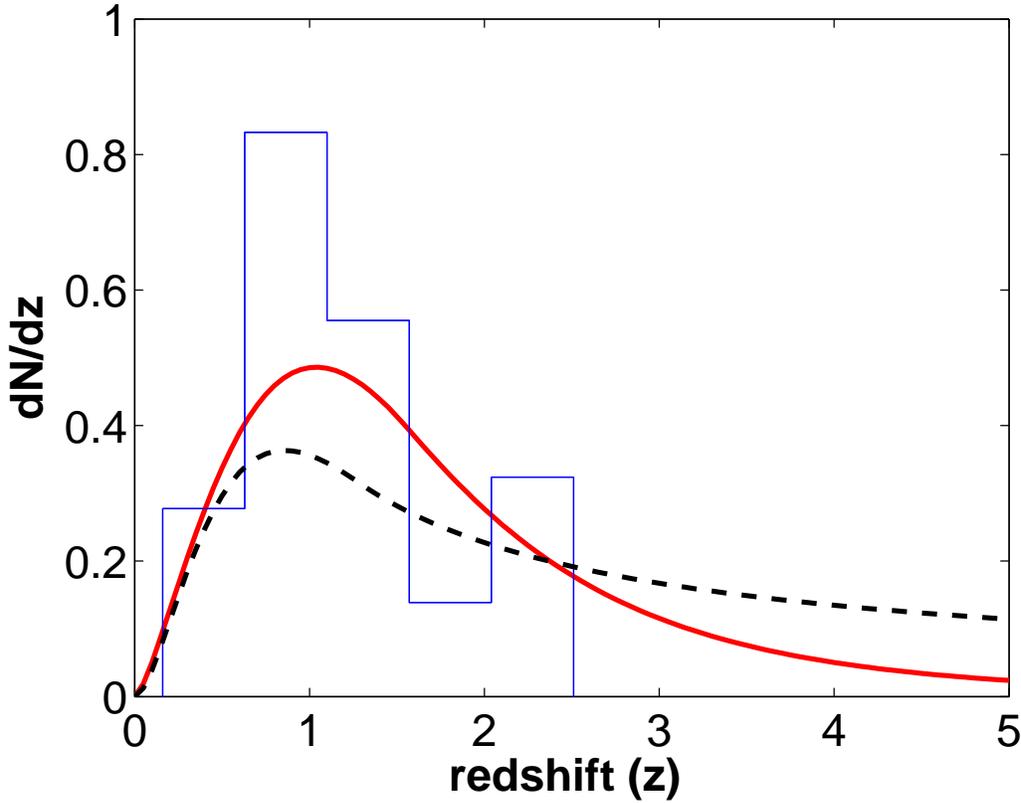} 
        	\caption{The histogram is constructed from our FSRQ source list. The thick solid line shows the distribution for the exponential model. The thick dashed line shows the distribution for the power law model.} 
\label{fig-fsrqdndztaubeta}
\end{figure}

The derived luminosity function parameter values are given in Table \ref{Tab:lumfn-param}. The source distribution ($dN/dz$) has been over plotted with the model distribution for both exponential and power law evolution model in Fig. \ref{fig-fsrqdndztaubeta}. For the model luminosity functions the $95\%$ upper limit of low luminosity end index is taken.  

\begin{table}
\begin{center}
\caption{Luminosity Function Parameters  }\label{Tab:lumfn-param}


 \begin{tabular}{ccc}
  \hline\noalign{\smallskip}
 & Exponential evolution function ($\tau$) & Power Law evolution function ($\beta$)  \\
\hline\noalign{\smallskip}
$L_B$ (in $10^{46}$\,erg\,s$^{-1}$)  & 1.7$^{+1.8}_{-0.4}$  & 5.8$^{+2.5}_{-0.9}$ \\ 
  $\alpha_2$  & 2.5$\pm$0.2  & 2.4$\pm$0.2  \\
$\alpha_1$ (95$\%$ upper limit)  & 1.1  & 0.9  \\
$\alpha_1$ (99$\%$ upper limit)  & 1.3  & 1.1  \\
  \noalign{\smallskip}\hline
\end{tabular}
\end{center}
\end{table}

\section{Luminosity function for BL Lacs}
\begin{figure}[h]
\begin{minipage}[t]{0.495\linewidth}
	\centering
\includegraphics[width=59mm,height=58mm]{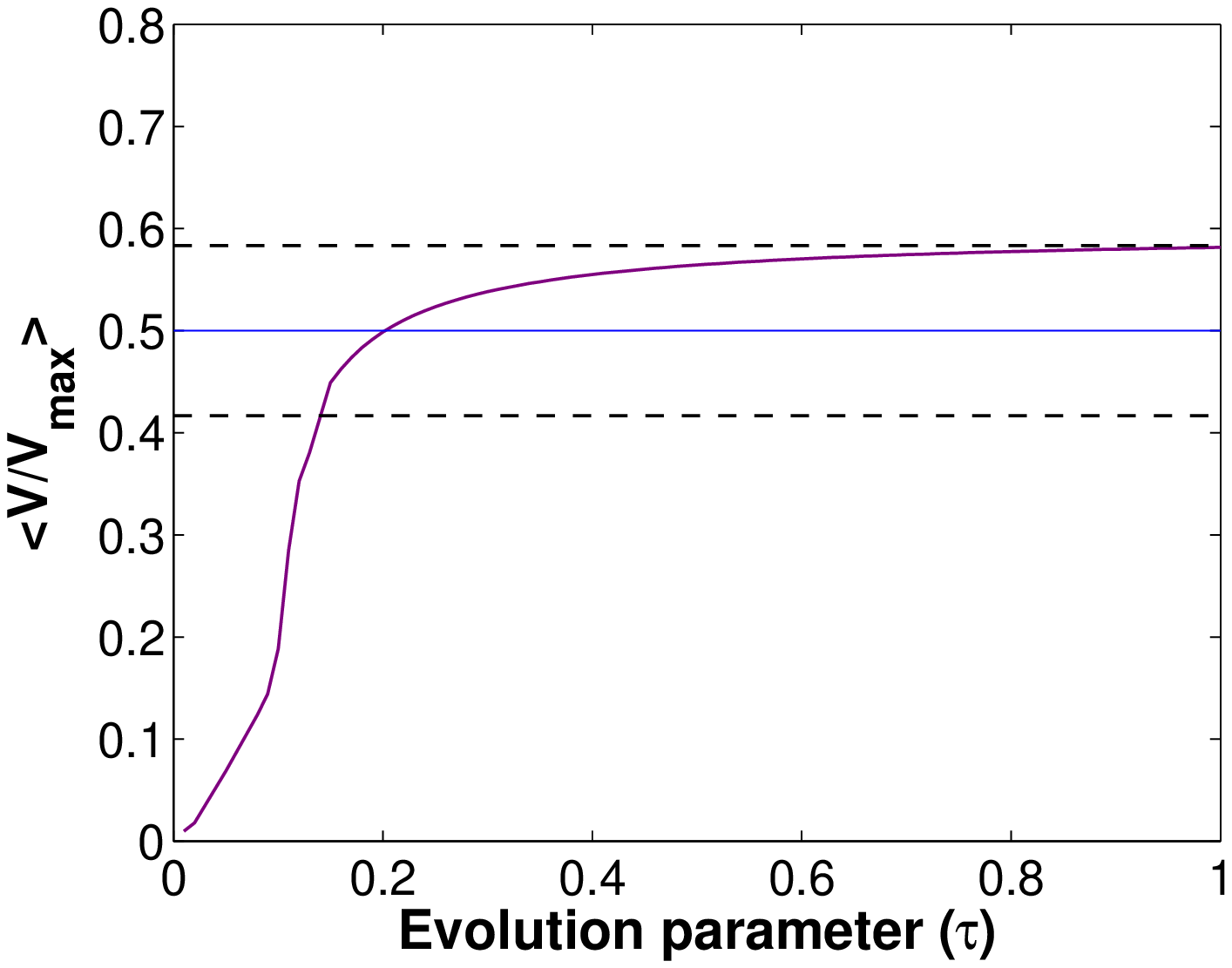}
	\caption{{\small Distribution of $<\frac{V}{V_{max}}>$ values for BL Lacs with different $\tau$ values. Dashed horizontal lines indicate the 1 $\sigma$ error in $<\frac{V}{V_{max}}>$.}} 
	\label{fig-tauvvmbl}
\end{minipage}
\begin{minipage}[t]{0.495\textwidth}
	\centering
\includegraphics[width=59mm,height=58mm]{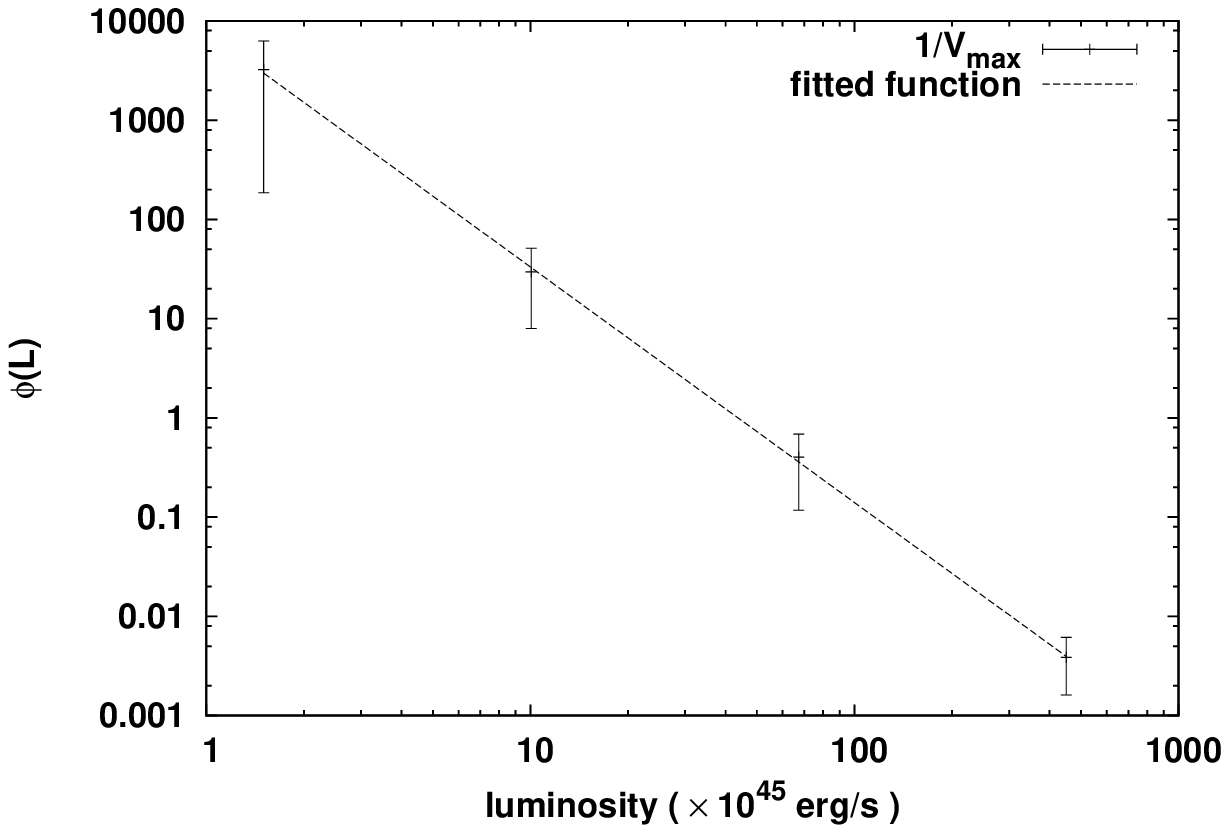}
	\caption{{\small $\phi(L)$ vs $L$ plot for BL Lacs ($\frac{1}{V_{max}}$ method). $\phi(L)$ is given in $(\frac{3}{4\pi})\times (\frac{H_0}{c})^{3}$ per luminosity (10$^{45}$ erg s$^{-1}$ unit) interval.}} 
	\label{fig-bllumfn}
\end{minipage}
\end{figure}

$<\frac{V}{V_{max}}>$ study shows no evidence of evolution for BL Lacs (Fig. \ref{fig-bllacvvm1}). We studied the variation of $<\frac{V}{V_{max}}>$ with the evolution parameter ($\tau$) of a pure luminosity evolution function (similar to FSRQs). It is found that $<\frac{V}{V_{max}}>$ is not sensitive to the $\tau$ variation (Fig. \ref{fig-tauvvmbl}). We construct the luminosity function of BL Lacs by following similar methods used in the case of FSRQs. We find a single power law luminosity function with an index of $(2.37 \pm 0.03)$ by $\frac{1}{V_{max}}$ method (Fig. \ref{fig-bllumfn}). 

Since the source list of BL Lacs is too small to independently determine the lower end luminosity index and break luminosity for BL Lacs, we assumed values similar to those derived for FSRQs.

\subsection{Average EGRET Limiting Flux}
The EGRET flux limit was not found to be identical for all sources. The empirical relationship between flux limit and statistical significance is given by (Chiang et al.~\cite{b6}), 
\begin{equation}
F_{i,lim} = F_{i} \frac{n_{lim}}{n_{i,sig}} 
\end{equation}
While calculating the $V_{max}$ of each source and also while finding the low end luminosity index and break luminosity, the appropriate flux limit for each source has been calculated. In order to compare the observed density distribution with the model distribution, the average limiting flux ($\sim 1\times 10^{-7}$\,ph\,cm$^{-1}$\,s$^{-1}$) is used. 

\section{Normalization constant of luminosity function}

The normalization of the luminosity function has been found by integrating the luminosity function over the redshift range zero and $z_{max}$, and above the EGRET limiting luminosity.
\begin{equation}
N_{obs} = \int_{0}^{z_{max}} \frac{dV}{dz}dz \int_{L_{lim}(z)}^{\infty} \phi(L_0)dL_0
\end{equation} 
Here, $N_{obs}$ is the number of FSRQs (BL Lacs) detected by EGRET. Since each EGRET source has a different limiting flux, in order to calculate the limiting luminosity $L_{lim}(z)$, the average limiting flux($F_{lim,ave}$) has been used ($1\times 10^{-7}$\,ph\,cm$^{-2}$\,s$^{-1}$). The limiting luminosity is given by
\begin{equation}
L_{lim}(z) = \frac{4\pi \times D_{L}^{2} F_{lim,ave}}{(1+z)^{2-\alpha} \times f(z)}
\end{equation}
Here $\alpha$ is the average photon spectral index of EGRET detected FSRQs (BL Lacs) and $f(z)$ is the $\gamma$-ray evolution function of FSRQs. For BL Lacs, $f(z)$ is unity.
Also since the coverage in the southern hemisphere is $0^{\circ} \le b \le -40^{\circ}$, a correction factor for the loss of solid angle is included.

\section{Discussion}
We investigated for the first time the luminosity function and evolutionary nature of FSRQs and BL Lacs separately in $\gamma$-rays. The construction of the luminosity function is of great importance in order to understand the nature of these populations as a whole. It also plays an important role in estimating their contribution to the $\gamma$-ray background. We find no evolution for BL Lacs, whereas for FSRQs, there is strong indication of evolution in $\gamma$-rays. Padovani et al.~\cite{padovani2007} found the luminosity function of BL Lacs and FSRQs using radio and X-ray data. They also found no evidence of evolution for BL Lacs, whereas FSRQs show strong evolution. For FSRQs, we assumed pure luminosity evolution. Two types of evolution functions (power law and exponential) are considered in this work. The source luminosities are de-evolved to zero redshift and the de-evolved luminosity function is constructed by $\frac{1}{V_{max}}$ method. This analysis only gives the high luminosity end power law index of the luminosity function. Assuming a broken power law model, we used likelihood analysis of source density distribution to find the break luminosity and low luminosity end index. Only an upper limit for the low end index is derived. For BL Lacs, the luminosity function is described by a single power law with inadequate data to derive a break luminosity. 
\par In earlier work, Chiang \& Mukherjee~(\cite{b5}) found a much flatter high end luminosity index ($\alpha_2$=2.2) than our estimation ($\alpha_2$=2.5 for exponential model and 2.4 for power law model). They also found a lower break luminosity ($L_B = 1.1\times 10^{46}$\,erg\,s$^{-1}$) than our estimation (given in Table \ref{Tab:lumfn-param}). Contrary to this work, they studied the FSRQs and BL Lacs as a single source class. We constructed the luminosity function and evolution of FSRQs and BL Lacs with a source list having almost twice the number of sources than Chiang \& Mukherjee~(\cite{b5}). We used a $\Lambda CDM$ cosmology, while they considered an Einstein-deSitter universe. Though we do not have any quantitative measure to select between the two forms of the evolution function, the source distribution (Fig. \ref{fig-fsrqdndztaubeta}) suggests that the exponential evolution function is a better representation than the power law evolution function. 
\par The low end luminosity function is not well constrained due to a lack of adequate number of sources. Considering the much higher sensitivity of FGST, one expects a clearer picture to emerge of the distribution of these source classes from FGST data. The blazar contribution to the Extragalactic Gamma-Ray Background (EGRB: see, Sreekumar et al.~\cite{sreekumar1998}; Strong, Moskalenko \& Reimer~\cite{strongegrb2004b}) together with the contribution from AGNs with different inclination angles, will be presented elsewhere.

\end{document}